\documentclass[aps,prsuperscriptaddress,twocolumn,reprint]{revtex4-1}
\usepackage{graphicx}
\usepackage{mathrsfs}
\usepackage{bm}
\usepackage{amsmath}
\usepackage{dcolumn}
\usepackage{epstopdf}
\usepackage{dsfont}
\usepackage{amssymb}
\usepackage{tabularx}
\usepackage{array}
\usepackage{color}
\usepackage[colorlinks=true, letterpaper=true, pdfstartview=FitV, linkcolor=blue, citecolor=blue, urlcolor=blue]{hyperref}

\begin{document}


\title{Intrinsic Nonreciprocity in Electron-Phonon Interaction Driven Thermoelectric Diodes}

\author{Hao-Kun Ke}
\affiliation{Department of Physics, School of Physics and Materials Science, Guangzhou University, Guangzhou 510006, China}

\author{Lie-Run Tian}
\affiliation{Department of Physics, School of Physics and Materials Science, Guangzhou University, Guangzhou 510006, China}

\author{Pei-Hao Fu}
\affiliation{Department of Physics, School of Physics and Materials Science, Guangzhou University, Guangzhou 510006, China}

\author{Jun-Feng Liu}
\email{phjfliu@gzhu.edu.cn}
\affiliation{Department of Physics, School of Physics and Materials Science, Guangzhou University, Guangzhou 510006, China}

\author{Jun Wang}
\email{jwang@seu.edu.cn}
\affiliation{Department of Physics, Southeast University, Nanjing 210096, China}

\author{H. Xu}
\email{xuh@sustech.edu.cn}
\affiliation{Department of Physics and Institute for Quantum Science and Engineering, Southern University of Science and Technology,
 Shenzhen 518055, China}

\begin{abstract}
	We study an electron-phonon interaction driven thermoelectric diode. The nonreciprocity in this diode arises from the asymmetry between the probabilities of phonon emission and absorption in the electron-phonon interaction, as well as the structural reflection asymmetry. We reveal the intrinsic nature of this nonreciprocity, as the forward and backward electron transport remains asymmetric even when the applied temperature difference is not reversed. This intrinsic nonreciprocity gives rise to two novel transport phenomena. One is a novel thermoelectric effect which is driven by the temperature difference between the leads and the central device region, rather than the conventional temperature difference between the two leads. The second, and more significant, phenomenon is the suppression of electronic backscattering in the load resistor. This suppression decreases the resistance of the load resistor, which leads to the breakdown of Ohm's addition law. Under suitable conditions, the presence of electron-phonon interaction can yield a larger thermoelectric current compared to the case without it. This intrinsic nonreciprocity opens up a new pathway for low-power electronics besides topology and superconductivity, and for nonreciprocal thermoelectric devices.
\end{abstract}

\maketitle

\textit{Introduction.---}
Nonreciprocal transport, characterized by a directional preference in charge or energy flow under reversed external stimuli, has become a cornerstone for rectification and energy harvesting in modern electronic systems \cite{2024GeneralTheoryforLongitudinal, 2024GiantnonlinearHallandwireless, 2023Thesuperconductingdiode,2024Colossalroomtemperature, 2024Highefficiencyenergyharvesting, 2022Rectificationof, 2025NonreciprocalCoulombdrag, 2024ScatteringTheoryof, 2024TransportandNonreciprocity, 2024NonreciprocalTransportandOpticalPhenomena, 2018Nonreciprocalresponsesfromnoncentrosymmetric}. While voltage-driven nonreciprocity -- exemplified by p-n junction diodes -- has been extensively studied, recent advances in thermoelectric diodes highlight temperature-gradient-driven nonreciprocal phenomena, such as thermal rectification \cite{2024ScatteringTheoryof, 2018ElectronTransferInducedThermal, 2020Thermalrectification, 2020Hightemperaturesiliconthermal, 2020Heatrectificationviaa, 2020Dualmodesolidstatethermal, 2022Simultaneouselectricalandthermal} and nonlinear Hall effects \cite{ 2025InverseSpinThermalHallEffect, 2024MicroscopictheoryofnonlinearHall, 2023Quantumkinetictheoryof, 2022GiantanomalousthermalHall, 2022Fundamentaldistinctionbetweenintrinsic, 2022ChiralanomalyinducednonlinearNernst} under thermal bias. Despite progress in characterizing these effects, a fundamental distinction remains unexplored: whether nonreciprocity is extrinsic (arising solely from stimulus reversal) or intrinsic (persisting even without reversing the temperature difference or voltage polarity). Current frameworks predominantly assume symmetry between forward and backward transport probabilities under fixed stimuli, overlooking the possibility of inherent asymmetry in microscopic scattering processes \cite{, 2011Scatteringmatrixapproach,2022NonlinearLandauerformula, 2005QuantumTransportAtom}. This constraint obscures the potential for fundamentally new transport regimes beyond conventional rectification mechanisms.

The Ohm's addition law of resistance, governing classical and quantum systems alike, imposes stringent limitations on energy-efficient electronics \cite{2022NonlinearLandauerformula,2005QuantumTransportAtom,2017ViolationofOhm,2012OhmsLawSurvives,2009ElectronicTransportontheNanoscale}. This law naturally emerges in incoherent quantum transport regimes, such as the phase-averaged limit \cite{sm1}. This law stems from the sequential tunneling in series barriers in the incoherent limit \cite{buttiker88} and has never been broken in the zero bias limit. Even in nearly dissipationless topological edge states or superconductors, the additive resistance of serially connected components remains valid in the phase-averaged incoherent limit, despite the resistivity is extremely small \cite{2023ComplexLandaulevelsandrelated, 2021Suppressingrelaxationthrough, 2008Influenceofdephasing, 2009TopologicalInsulatorANewQuantized, 1999HallResistivityandDephasing, 2011Dissipativechargetransportindiffusive}. This universality underscores the dominance of Ohmic dissipation in long-range conduction. However, the emergence of intrinsic nonreciprocity challenges this paradigm: if electronic backscattering becomes directionally suppressed, the resistance of downstream components could deviate from classical summation. Such a breakdown would not only redefine resistive networks but also unlock low-dissipation pathways -- a transformative prospect that remains unexplored.

Here, we establish intrinsic nonreciprocity in an electron-phonon (e-ph) interaction driven thermoelectric diode, where asymmetry in phonon emission/absorption probabilities \cite{2024Coherenceenhanced, 2019Quantumdotcircuita, 2015Phononthermoelectrictransistors, 2014TheNonEquilibriumGreens, 2008ModelingofNanoscaleDevices} and structural reflection symmetry breaking jointly induce directionally locked transport. Crucially, forward and backward scattering probabilities remain asymmetric under a fixed temperature difference, defying extrinsic rectification mechanisms. This intrinsic property manifests in two groundbreaking phenomena: (1) A thermoelectric current driven not by the conventional inter-lead temperature difference but by the temperature difference between the leads and the central device region, redefining the thermodynamic driving force; (2) Suppression of electronic backscattering in the downstream resistor, violating Ohm's additive resistance law even in the phase-averaged incoherent transport. Strikingly, e-ph interactions —- conventionally viewed as a source of dissipation —- enhance the thermoelectric current under specific conditions, overturning the common belief of phonon-induced performance degradation. Our work unveils a paradigm for low-power electronics rooted in intrinsic nonreciprocity, complementing topological and superconducting approaches while operating at higher temperatures and zero magnetic fields.

\begin{figure}[t]
	\begin{center}
		\includegraphics[width=3.415in]{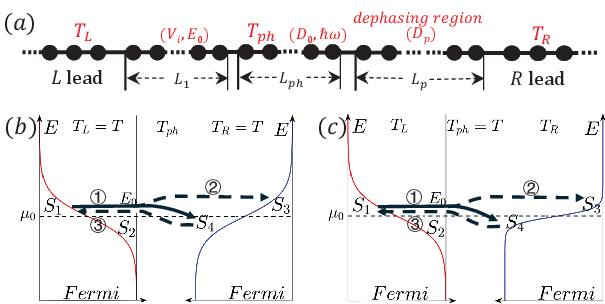}
	\end{center}
	\caption{(a) Device Schematic: A nanowire (NW) is connected to the $L$ and $R$ leads, comprising three regions: a tunneling junction (region 1, length $L_1$) with a resonance energy level $E_0$; an e-ph interaction region (region 2, length $L_{ph}$) characterized by temperature $T_{ph}$, scattering intensity $D_0$, phonon energy $\hbar\omega$; an elastic dephasing resistor (region 3, length $L_p$) with scattering intensity $D_p$ of zero-energy phonons. (b-c) Schematic diagrams of electron transport processes with e-ph interaction for novel and conventional thermoelectric effect, repectively. The average temperature of the $L$ and $R$ leads is $T = (T_L + T_R)/2$. In (b), $T_L = T_R = T$ and $\Delta T = |T-T_{ph}|$. In (c), $T_{ph} = T$ and $\delta T = |T_L-T_R|$.  Electrons exhibit predominant conduction at the resonance energy $E_0$, while the transport is significantly suppressed in off-resonance energy ranges (as indicated by the black lines in the sketchs)‌.
 }
	\label{Fig1}
\end{figure}

\textit{Model.---}
We consider a nanowire (NW) modeled by a one-dimensional lattice (see Fig. \ref{Fig1}(a)) connected to two leads $(L, R)$, partitioned into three distinct regions. The first segment with length $L_{1}$, region 1, is a tunneling junction with resonant energy level $E_0$. The second segment with length $L_{ph}$, region 2, is the region including inelastic e-ph interaction. The third segment with length $L_p$, region 3, acts as a dephasing resistor including dephasing processes modeled by also e-ph interaction with zero-energy phonons. The whole scattering region is $i\in[1,L]$ with $L=L_1+L_{ph}+L_p$. The entire device is subjected to a temperature gradient, where $T_{L}$, $T_{R}$, and $T_{ph}$  denote the temperatures of the left lead, right lead, and the central scattering region. To independently tune $T_{ph}$, the central scattering region may be coupled to a phonon bath that exchanges phonons —- but not electrons —- with the device \cite{prb14,prb10,rmp06,rpp75}.

The tight-binding Hamiltonian for the NW can be written as
\begin{equation}
	H=\sum_{i}t(c_{i}^{\dagger}c_{i+1}+h.c.) + \sum_{i}U_{i}c_{i}^{\dagger}c_{i}.
	\label{Eq1}	
\end{equation}
Here $c_{i}^{\dagger}$ creates an electron at site $i$, $t =-1eV$ is the hopping energy in the wire. The on-site energy $U_{i}=20t$ are applied only in region 1 for the sites $i=1,3,4,6$ and $U_{i}=0$ for other sites. The sites within region 1 are numbered sequentially from left to right as $1, 2, ..., L_1$. This on-site potential configuration constructs a tunneling junction with a resonance energy level $E_0 \approx 100$ meV and a broadening about $7$ meV in region 1.

To study the current response under a temperature bias, we calculate the retarded Green’s function of the system. The effect of e-ph interaction in regions 2 and 3 is incorporated as an interaction self-energy term for electrons $\Sigma_{ph}$ \cite{2005QuantumTransportAtom, 2014TheNonEquilibriumGreens, 2008ModelingofNanoscaleDevices}, and self-energy of the dephasing process $\Sigma_{s}$ \cite{2005QuantumTransportAtom, 2008ModelingofNanoscaleDevices, 2007NonequilibriumGreens}, respectively. By means of the lattice Green's function technique \cite{sm2}, we calculate the current spectra $J_{L,R}(E)$ and integrate it over energy to obtain the total current $J_{L,R}$ for two leads. For current conservation, we have $J_L=-J_R$. We set the current through the device as $J=J_L=-J_R$.

We are interested in thermoelectric transport driven only by a temperature difference. The two leads have the same chemical potentia $\mu_L = \mu_R = \mu_0
= 90 meV$. The charge current in a temperature configuration is denoted as $J(T_L,T_{ph},T_R)$, and the average temperature of two leads is defined as $T = (T_L + T_R)/2$. For a conventional thermoelectric effect, a positive or negative temperature bias with the magnitude $\delta T = |T_L-T_R|$ between two leads drives
a charge current $J^{+}(T+\delta T/2, T,T-\delta T/2)$ or $J^{-}(T-\delta T/2,T,T+\delta T/2)$. Note that $T_{ph}=T$ should be satisfied to ensure a convergent, well-defined thermopower in nonreciprocal cases. Even at zero $\delta T$, a positive or negtive temperature difference with the magnitude $\Delta T = |T-T_{ph}|$ also drives a charge current
$J^{+}(T, T-\Delta T, T)$ or $J^{-}(T, T+\Delta T, T)$. This is a novel thermoelectric effect unique to a thermoelectric diode with intrinsic nonreciprocity. The nonreciprocal efficiency for the novel (conventional) thermoelectric effect $\eta^{N(C)}$, is given by
\begin{equation}
	\eta^{N(C)} = \frac{J^{+}+J^{-}}{|J^{+}|+|J^{-}|}.
	\label{Eq6}	
\end{equation}

\begin{figure}[t]
	\begin{center}
		\includegraphics[width=3.415in]{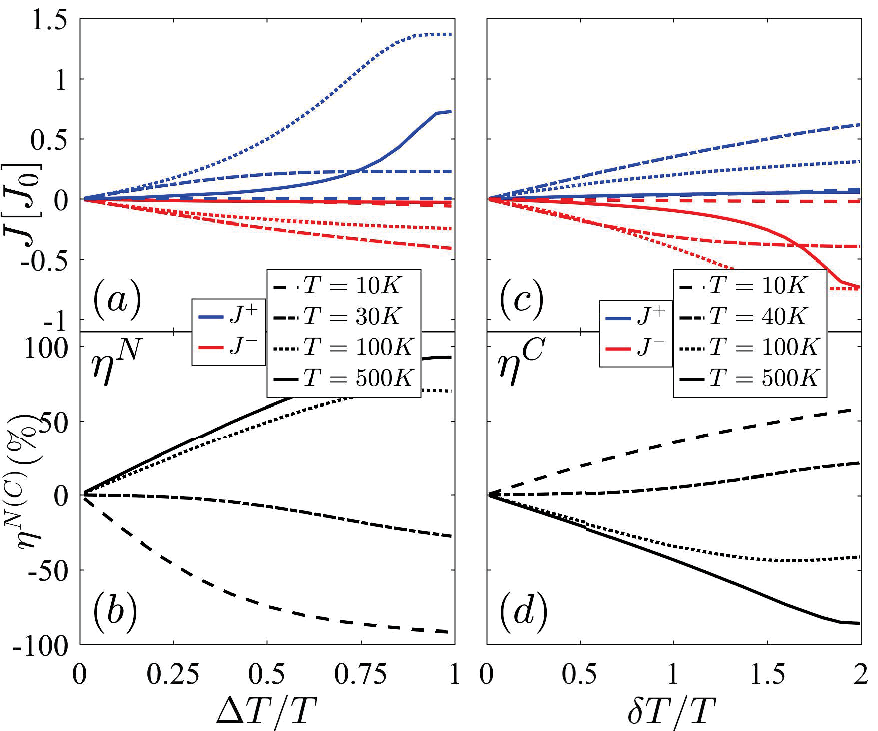}
	\end{center}
	\caption{The charge currents $J^{\pm}$ and the nonreciprocal efficiencies $\eta^{N}$ or $\eta^{C}$ as functions of the temperature difference $\Delta T$ or $\delta T$ for the novel (a) and (b) or conventional (c) and (d) thermoelectric effect. Various average temperatures $T$ are considered. The other parameters are: $L_{ph} = 10$, $L_1 = 6$, $\hbar \omega = 5$ meV and $D_0 = 1$. $J_0 = 10^{-3} e^2V/h$.}
	\label{Fig2}
\end{figure}

\begin{figure*}[t]
	\begin{center}
		\includegraphics[width=5.8in]{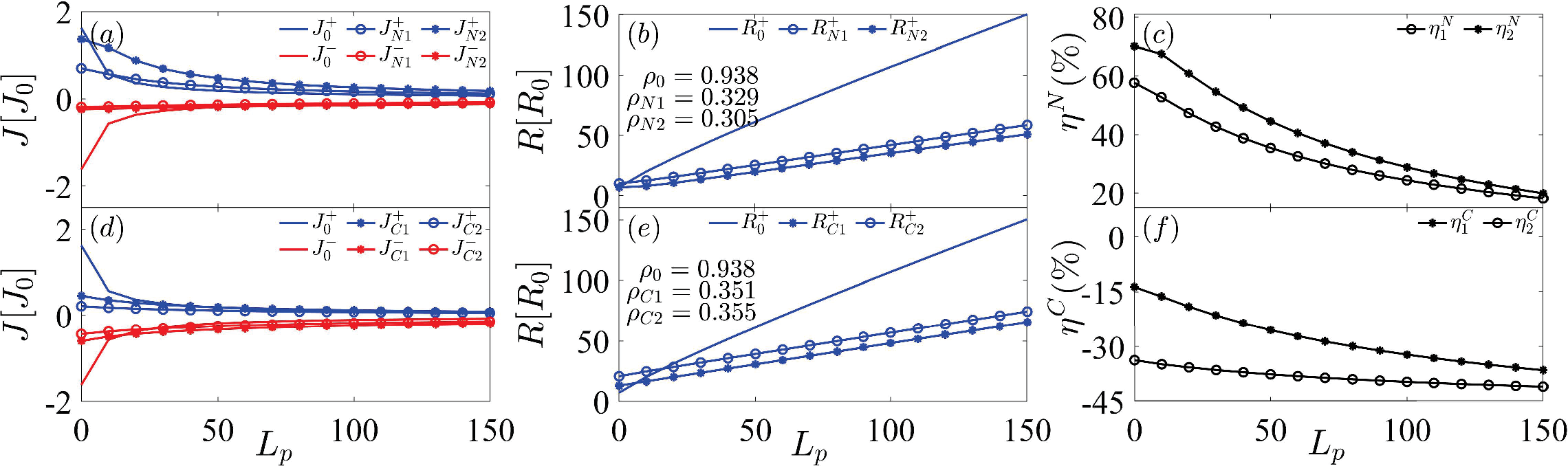}
	\end{center}
	\caption{The charge current, the resistance, and the nonreciprocal efficiency as functions of $L_p$ for the novel thermoelectric effect (a)-(c), and the conventional thermoelectric effect (d)-(f), respectively. Systems $N_1$, $N_2$ ($C_1$, $C_2$), and system $0$ are considered in the novel (conventional) thermoelectric effect. In system $0$, $\delta T = 100 K$, $\Delta T = 0$, $D_0 = 0$. In system $N_1$ ($N_2$), $\Delta T = 60K$ ($100K$), $\delta T = 0$, $D_0 = 1$. In system $C_1$ ($C_2$), $T = 60K$ ($100K$), $\delta T = 100K$, $\Delta T = 0$, $D_0 =1$. $D_0 = D_p=1$. The other parameters are: $L_{ph} = 10$, $L_1 = 6$, $\hbar \omega = 5$ meV, $D_p = 1$, $R_0 = h/e^2$.} 	
	\label{Fig3}
\end{figure*}

\textit{Novel nonreciprocal thermoelectric effect.---}
The intrinsic nonreciprocity in the model arises from the combination of the asymmetry between the probabilities of phonon emission and absorption in the e-ph interaction and the structural reflection asymmetry. At equilibrium, the number of phonons is $N_{\omega} = 1/\{\exp[\hbar\omega / (k_B T_{ph})]-1\}$. The phonon absorption rate $K^{ab}$ is proportional to $N_{\omega}$, while the emission rate $K^{em}$ is proportional to $N_{\omega} + 1$ \cite{2005QuantumTransportAtom}. The ratio of emission to absorption is $K^{em}/K^{ab} = (N_{\omega}+1)/N_{\omega} = \exp[\hbar\omega/ (k_B T_{ph})]$. This asymmetry between phonon emission and absorption dramatically depends on the temperature. At low temperature, electrons prefer to emit phonons while the absorption is heavily suppressed.

We first consider the novel thermoelectric effect with $T_L=T_R=T$, $\Delta T\neq0$, as shown in Fig. \ref{Fig1}(b). The energy range of the left (right) lead is divided into two regions, $S_1$ ($S_3$) region for $E>\mu_0$ and $S_2$ ($S_4$) region for $E<\mu_0$. Electrons from $S_1$ are filtered by the resonance tunneling junction in region 1. Then electrons with energy around $E_0$ can emit (absorb) phonons and flow through region 2, finally enter into $S_4$ ($S_3$), as depicted by process $\textcircled{1}$ ($\textcircled{2}$). The reverse process of $\textcircled{1}$ from $S_4$ to $S_1$ via phonon absorption is labeled as process $\textcircled{3}$. In the zero temperature limit $T_{ph}=0$, only process $\textcircled{1}$ survives and contributes a positive charge current, which is a novel thermoelectric effect. With increasing $T_{ph}$, the positive charge current decreases due to the emergence of processes $\textcircled{3}$. When $T_{ph}=T$, processes $\textcircled{1}$ and $\textcircled{2}$ reach equilibrium with processes $\textcircled{3}$, resulting in no net current. Although the probability of phonon emission is still higher than that of phonon absorption, the electron occupation in $S_4$ is larger than in $S_1$. For higher temperature $T_{ph}>T$, the ratio of phonon emission to absorption decreases, thus process $\textcircled{3}$ prevails over process $\textcircled{1}$ and $\textcircled{2}$, resulting in a net negative current.

Fig. \ref{Fig2}(a) and (b) display the charge current and the nonreciprocal efficiency as functions of the temperature difference for the novel thermoelectric effect with various lead temperatures. $D_{0}$ represents the e-ph scattering strength. Several key features are noteworthy: (i) the magnitude of the charge current increases with increasing temperature difference; (ii) the nonreciprocal efficiency at small $\Delta T$ is positive (negative) for high (low) lead temperature $T$, the critical temperature almost $T=30 K$ is determined by the difference of the resonance level and the Fermi energy $E_0-\mu_0$. The magnitude of the nonreciprocal efficiency can reach above $90\%$ at large $\Delta T$. This extrinsic nonreciprocity stems from the nonlinear dependence of emission/absorption ratio $K^{em}/K^{ab}$ on $T_{ph}$. The thermoelectric current and nonreciprocity as functions of the e-ph interaction strength are also presented in the Supplemental Material (SM) \cite{sm3_Fig5}, which reveals that the e-ph interaction plays the key role in the novel thermoelectric effect. Although this novel thermoelectric effect has been described in literatures \cite{prb10}, the nonreciprocity has not been investigated.

\textit{Breakdown of Ohm's addition law.---} To further manifest the intrinsic nonreciprocity in the thermoelectric effect, we show the suppression of downstream resistance, which breaks Ohm's addition law. As shown in Fig. \ref{Fig1}(a), the downstream resistor is modeled by a dephasing region of length $L_p$. During the transport in the region, both the phase and the momentum of electrons relax simultaneously by the e-ph scattering with scattering intensity $D_p$ of zero-energy phonons. In the dephasing, or phase average limit, the usual phase-induced mesoscopic phenomena, such as conductance fluctuation and Anderson localization, are suppressed. The Ohm's addition law of resistors applies and the resistivity is well-defined. We will demonstrate the suppression of the resistivity of loaded resistor due to the intrinsic nonreciprocity.

Specifically, we consider three configurations of parameters, labelled by system $0$, $N_1$, and $N_2$, with the same average temperature $T = 100 K$. In system $0$, the temperature difference between two leads $\delta T = 100 K$,  $\Delta T = 0$, $D_0 = 0$. A conventional thermoelectric effect without e-ph interaction in region 2 is considered as a comparison. The novel thermoelectric effect is considered in system $N_1$ and $N_2$. In system $N_1$ ($N_2$), $\Delta T = 60K$ ($100K$), $\delta T = 0$, $D_0 = 1$. Fig. \ref{Fig3}(a-c) show the charge current, the resistance, and the nonreciprocal efficiency as functions of $L_p$ for three systems, respectively. $D_p$ represents the dephasing strength. The resistance of the device is defined via an effective thermoelectric voltage \cite{sm5}. With fixed temperature difference, the thermoelectric voltage is assumed to be unchanged with increasing $L_p$. In system $0$, the current is antisymmetric with respect to the temperature difference, $|J^+_0| = |J^-_0|$. The nonreciprocity is absent. In contrast, in system $N_1$ and $N_2$ with e-ph interaction, $|J^+_{N1(N2)}|>|J^-_{N1(N2)}|$ due to the intrinsic rightward nonreciprocity, which leads to a large positive nonreciprocal efficiency shown in Fig. \ref{Fig3}(c). With increasing $L_p$, both the charge current and the nonreciprocal efficiency decrease. The decreasing of current can be attributed to the increasing resistance of region 3.

The total resistance defined via the effective thermoelectric voltage is shown in Fig. \ref{Fig3}(b) as the function of $L_p$ when $T>T_{ph}$. Then we define the resistivity $\rho$ of region 3 as the slope obtained from a linear fit of the resistance versus $L_p$. It is shown that the resistivity $\rho_{N_1} (\rho_{N_2})<< \rho_0$ in system $N_1$ ($N_2$), indicating that the strong intrinsic nonreciprocity leads to the strong suppression of the downstream resistance. This is because the left-moving electrons backscattered in region 3 are reflected by regions 1 and 2 into right-moving electrons due to the suppression of process $\textcircled{3}$. The same resistor, region 3, has different resistance, in system $0$, $N_1$ and $N_2$. It implies the breakdown of Ohm's addition law of serially connected resistors.

In the simple case where electrons flow through a single energy channel, more understanding and details are provided in the SM \cite{sm1} about how the intrinsic nonreciprocity decreases the downstream resistance in the incoherent limit. Even in the transport with inelastic scattering, the concepts and qualitative results still hold. But the suppression effect of the downstream resistance is weaken due to the average of multiple energy channels.

\textit{Conventional nonreciprocal thermoelectric effect .---}
Next, we consider the conventional thermoelectric effect driven by $\delta T$ instead of $\Delta T$. The temperature of region 2 is set to be equal to the average temperature of two leads, i.e., $T_{ph}=T$. When $T_L = T_R$, processes $\textcircled{1}$ and $\textcircled{2}$ reach equilibrium with process $\textcircled{3}$, resulting in zero net current. When $\delta T\neq0$, $T_L=T+\delta T/2$, and $T_R=T-\delta T/2$, the electron populations in both $S_1$ and $S_4$ regions increase, exhibiting identical magnitudes under the first-order approximation with respect to
$\delta T$. However, the phonon emission probability remains higher than that of absorption. Consequently, the current contribution from processes
$\textcircled{1}$ and $\textcircled{2}$ dominates over process $\textcircled{3}$, resulting in a net positive current. Reversely, when $T_L=T-\delta T/2$ and $T_R=T+\delta T/2$, the electron populations in both $S_1$ and $S_4$ regions decrease by the same magnitude. Similarly, due to $K^{em}/K^{ab}>1$, the current contribution from process $\textcircled{3}$ dominates over processes $\textcircled{1}$ and $\textcircled{2}$, resulting in a net negative current. The nonlinear dependence of the Fermi distribution function on the temperature gives rise to the extrinsic nonreciprocity of this conventional thermoelectric effect with respect to $\delta T$.

In Fig. \ref{Fig2}(c) and (d), the charge current and the nonreciprocal efficiency versus $\delta T$ with virous $T$ are presented. The nonreciprocal efficiency at small $\Delta T$ is positive (negative) for low (high) average temperature $T$, the critical temperature almost $T=40 K$ is determined by the difference of the resonance level and the Fermi energy $E_0-\mu_0$. The magnitude of the nonreciprocal efficiency can reach nearly $90\%$ when both $T$ and $\Delta T$ are large.

Analogous to the novel thermoelectric effect, we also investigate the suppression of downstream resistance due to the intrinsic nonreciprocity in this conventional thermoelectric effect. Specifically, we consider two configurations of parameters, labelled by system $C_1$ and $C_2$, in comparison with system $0$ mentioned above. In system $C_1$ ($C_2$), $T = 60K$ ($100K$), $\delta T = 100K$, $\Delta T = 0$, $D_0 = 1$.

Fig. \ref{Fig3}(d-f) show the charge current, the resistance, and the nonreciprocal efficiency as functions of $L_p$ for three systems $0$, $C_1$, and $C_2$, respectively. From the total resistance for $T_L>T_R$ shown in Fig. \ref{Fig3}(e), it is seen that the resistivity of region 3 $\rho_{C_1} (\rho_{C_2})<< \rho_0$ in system $C_1$ ($C_2$), exhibiting the strong suppression of the downstream resistance by the intrinsic nonreciprocity. Surprisingly, in contrast to the suppression of $\eta^N$ by increasing $L_p$, $|\eta^C|$ increases with increasing $L_p$.

More significantly, due to suppression of the load resistance, system $C_1$ generates larger current $|J_{C_1}^-|$ than system $0$ under the same $\delta T$ for $L_p>20$, as shown in Fig. \ref{Fig3}(d). This suppression under negative temperature differences ($T_L<T_R$) does not contradict the conclusion of downstream resistance suppression. Although the net flow is leftward, the actual resistance suppression occurs in process $\textcircled{1}$. When electrons in process $\textcircled{1}$ experience reduced load resistance and achieve higher transmission, the system requires self-consistent adjustment of the phonon emission-absorption ratio to alter the current in process $\textcircled{3}$. This adjustment rebalances processes $\textcircled{3}$ with $\textcircled{1}$ and $\textcircled{2}$ at zero temperature difference.
In this sense, the "upstream" resistance can also be suppressed.

Finally, we briefly outline the experimental feasibility of our proposed thermoelectric diodes. To detect the novel thermoelectric effect, the temperature difference $\Delta T = |T-T_{ph}|$ can be established by thermally anchoring the central e-ph region to an independently tunable phonon bath (e.g., via a microscale heater/cooler). The tunneling junction can be realized by a quantum point contact with gate-tuned potential profile. The difference between the resonance energy level of the quantum point contact and the chemical potential should be of the same order of magnitude as the dominant phonon energy in the e-ph region. The dephasing region naturally corresponds to a semiconductor nanowire segment with length exceeding its phase coherence length, as routinely achieved in disordered nanowires. Thus, the intrinsic nonreciprocity in the novel and conventional thermoelectric effects should be experimentally observable.

\textit{Conclusion.---}
We propose an e-ph interaction driven thermoelectric diode where the e-ph region can exchange energy with the environment. In this thermoelectric diode, we establish intrinsic nonreciprocity in that the forward and backward electron transport remains asymmetric even when the applied temperature difference is not reversed. This intrinsic nonreciprocity arises from the asymmetry between the probabilities of phonon emission and absorption in the e-ph interaction, as well as the structural reflection asymmetry. Beyond extrinsic nonreciprocal thermoelectric effects, this intrinsic nonreciprocity may also give rise to the novel thermoelectric effect and the load resistance suppression. Under optimized conditions, e-ph coupling enhances--rather than dissipates--current generation, overturning phonons' traditional dissipative role. The mechanism operates without topological constraints or superconductivity, offering advantages including room-temperature functionality and zero-field operation.

\section*{Acknowledgements}
The work described in this paper is supported by the National Natural Science Foundation of China (Grants No. 12174077 and No. 12174051).

\end{document}